\documentclass[floatfix,preprintnumbers,reprint,amsmath,amssymb,article]{revtex4}
\usepackage{graphics}
\usepackage{graphicx}
\usepackage{dcolumn}
\usepackage{bm}
\usepackage{multirow}
\usepackage{amssymb}
\usepackage{perpage}	 

\renewcommand\thesection{\arabic{section}}

\begin{document}

\title{Charge Pumping Through a Single Donor Atom}
\author{G. C. Tettamanzi$^{1}$}
\email{g.tettamanzi@unsw.edu.au}
\author{R. Wacquez$^{2}$}
\author{S. Rogge$^{1}$}
\affiliation{$^{1}$School of Physics and Australian Centre of Excellence for Quantum Computation and Communication Technology, UNSW, Sydney, Australia}
\affiliation{$^{2}$CEA, LETI, MINATEC Campus, 17 rue des Martyrs, 38054 Grenoble, France}

\begin{abstract} 
Presented in this paper is a proof-of-concept for a new approach to single electron pumping based on a Single Atom Transistor (SAT). By charge pumping electrons through an isolated dopant atom in silicon, precise currents of up to 160 pA at 1 GHz are generated, even if operating at 4.2 K, with no magnetic field applied, and only when one barrier is addressed by sinusoidal voltage cycles.
\end{abstract}

\maketitle

\date{\today}

\tableofcontents

\section{$\textbf{Introduction}$}
\label{sec:S1}

Although some fascinating alternatives have been recently suggested, e.g. semiconductor quantized voltage sources see Ref.~\cite{Hoh056802}, the main task of the electrical branch of quantum metrology remains the development of reliable quantised electron pumps (QEPs) \cite{Mil279,Gib1,Pot249,Kou1626,Kel1804,Blu343,Pek120,Jeh021012,Fuj042102,Kae012106,Sie3841} and, as a consequence, the full implementation of the quantum metrology triangle \cite{Mil279}. The QEP approach to metrology was introduced in the early 90s \cite{Pot249} and, since then, many interesting schemes have been proposed, each with advantages and disadvantages~--~see Ref. \cite{Gib1,Pot249,Kel1804,Blu343,Pek120,Fuj042102,Kae012106,Sie3841}. The strong focus on QEP-based metrology is also justified by its ability to generate currents in which capture and emission processes can be controlled at the single electron level. This can be obtained by taking advantage of Coulomb Blockade (CB) effects \cite{Bee1646} as QEPs are often Single Electron Transistor (SET) based \cite{Gib1,Pot249,Kel1804,Blu343,Pek120,Fuj042102,Kae012106,Sie3841}.

\begin{figure*}
\begin{center}
\includegraphics[width=172mm]{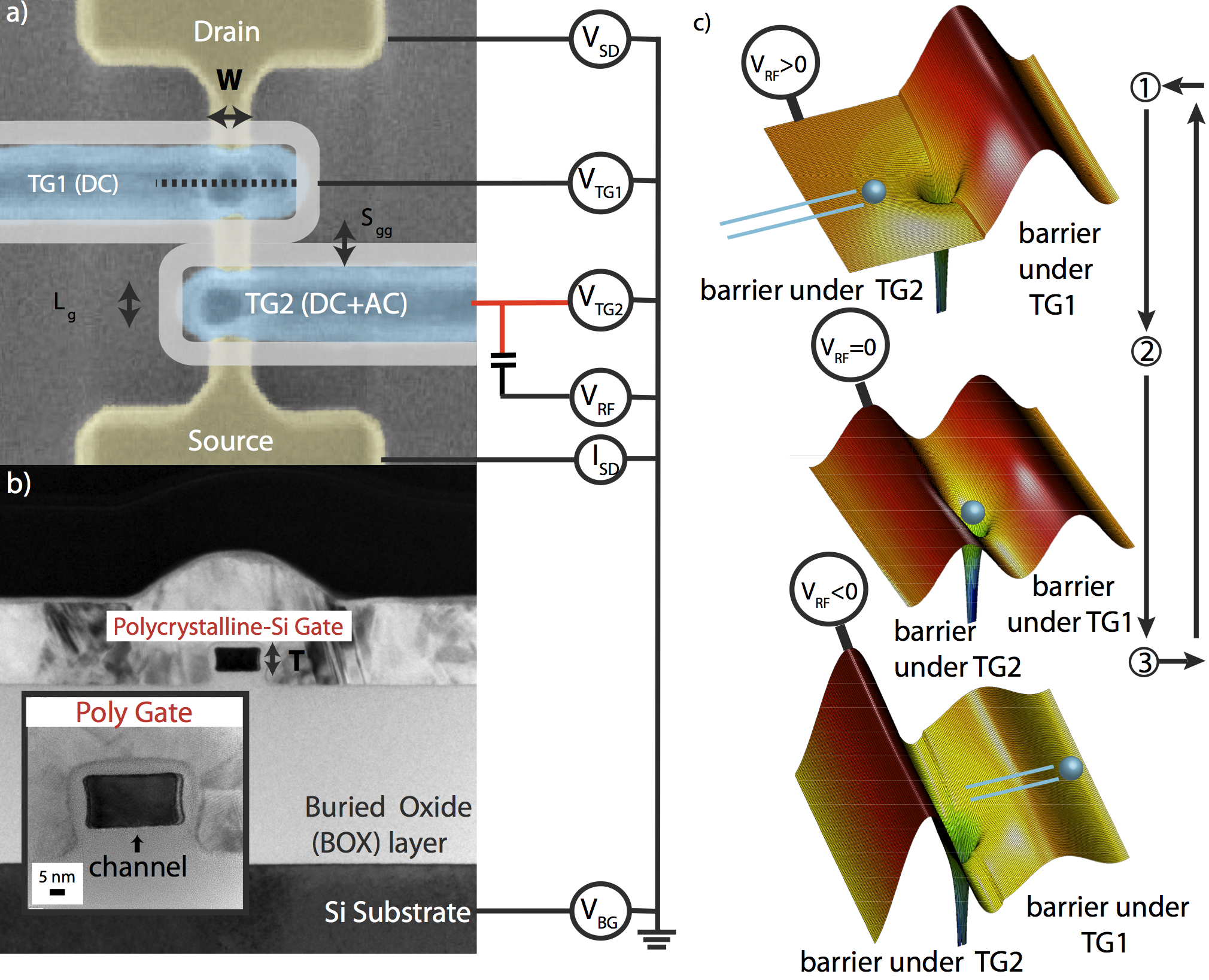}
\caption{a) Scanning electron micrograph and b) high resolution transmission electron micrograph (HR-TEM) of devices similar to the ones studied in this article. In a) the channel, the top gates at a distance $S_{gg}$ from each other and the spacers are shown in yellow, blue and light grey respectively. The HR-TEM, presented along the plane normal to the direction of transport, is indicated with the black dotted line in section a). This illustrates the structure of the channel, of the top gates surrounding it and of the substrate through which DC back gate voltages can be applied. The electrical circuit used in our experiments is standard \cite{Fuj042102,Kae012106} and schematically shown on the right of these sections a) and b). The cryogenic bias-T \cite{Gib1,Kae012106} is shown in red. c) Schematic diagrams illustrating the evolution of the potential landscape during each pumping cycle. In the first half of each cycle (1), the RF voltage applied to TG2 lowers the source-to-channel barrier and allows the electrons to be captured by the atomic potential. In (2), the electron is briefly bound to the atom, and in (3), when a sufficiently negative RF voltage is applied to the barrier in TG2, the electron escapes to the drain \cite{Gib1,Kae012106}. The system is then re-initialised for the next cycle. For device $A$, the RF voltage is added \textit{only} to TG2. The symmetrical situation of applying RF has been used for device $B$.}
\label{fg:PUMPING1}
\end{center}
\end{figure*}

To be useful for metrology, a QEP needs to generate currents of the order of hundreds of $pA$ with accuracies of 1 part in $10^{8}$ (equivalent to $10^{-2}$ ppm). Furthermore, as the current ($I$) is linked to the frequency of oscillation ($f$) according to the simple relation $I_{SD}=fe$ \cite{Pot249}, with $e$ being the elementary charge, $f$~$\gtrsim$~GHz is necessary to generate the desired currents. At GHz frequencies, QEPs are known to be affected by non-adiabatic excitation errors~\cite{Kat126801,Kas186805}. It is already understood that, to improve robustness and, as a consequence, temperature and speed of operations, their charging energy ($E_{C}$) needs to be increased~\cite{FleR16291}. The $E_{C}$ is the energy that needs to be paid to a system to increase the average number of electrons it contains by one \cite{Pot249,Blu343} and, for Quantum Dots (QDs), is inversely proportional to its size, which makes increasing it in conventional QD-SETs is a non-trivial task \cite{Shi103101}. Between the many approaches available, the one that has proven to be both successful and reproducible is based on single atom transistors (SATs)~\cite{Sel206805,Roc206812,Mor345,Fue242,Pla541,Tet046803}. 

High currents and high accuracies have been recently achieved \cite{Gib1} in a single-parameter charge pumping configuration \cite{Fuj042102,Kae012106} similar to the one used in this work and every possible novelty in the technology used to fabricate and to operate a QEP represents an important milestone \cite{Mil279,Hoh056802}. As this article will show, an isolated shallow dopant atom, due to the combination of a high $E_{C}$ and a well isolated ground state for the first electron charge state, provides, \textit{naturally}, an interesting geometry for the quantum pumping of electrons. This is not the first time that dopant atoms have been used to generate pumping currents. Lansbergen \textit{et al} \cite{Lan763} and more recently Roche \textit{et al} \cite{Roc1581} used a few dopants FET to perform pumping at MHz frequency, however, these pioneering experiments were not at the single atom level and far from the rates requested for quantum metrology. 

After a short introduction of the devices structure, of the selection process for the location of the good devices amongst the large pool of fabricated ones and of the experimental setup in section~\ref{sec:S2}, in section~\ref{sec:S3} we discuss the generation of pumped current up to the GHz frequencies and we show that, for our system, it is possible to reach the $I_{SD}$ = fe quantisation even for $f$~=~1~GHz. Some considerations on the consequences of our results are discussed in section~\ref{sec:S4}. In the appendix (section~\ref{sec:A1}) we give more insight on the differences between the non-adiabatic regime (studied in our experiment) and the adiabatic pumping regime. Lastly, more details on some aspects of our experiments are discussed in the appendix (sections~\ref{sec:A2},~\ref{sec:A3} and~\ref{sec:A4}).

\section{$\textbf{Devices selection and measurement setup}$}
\label{sec:S2}

\begin{figure*}
\begin{center}
\includegraphics[width=172mm]{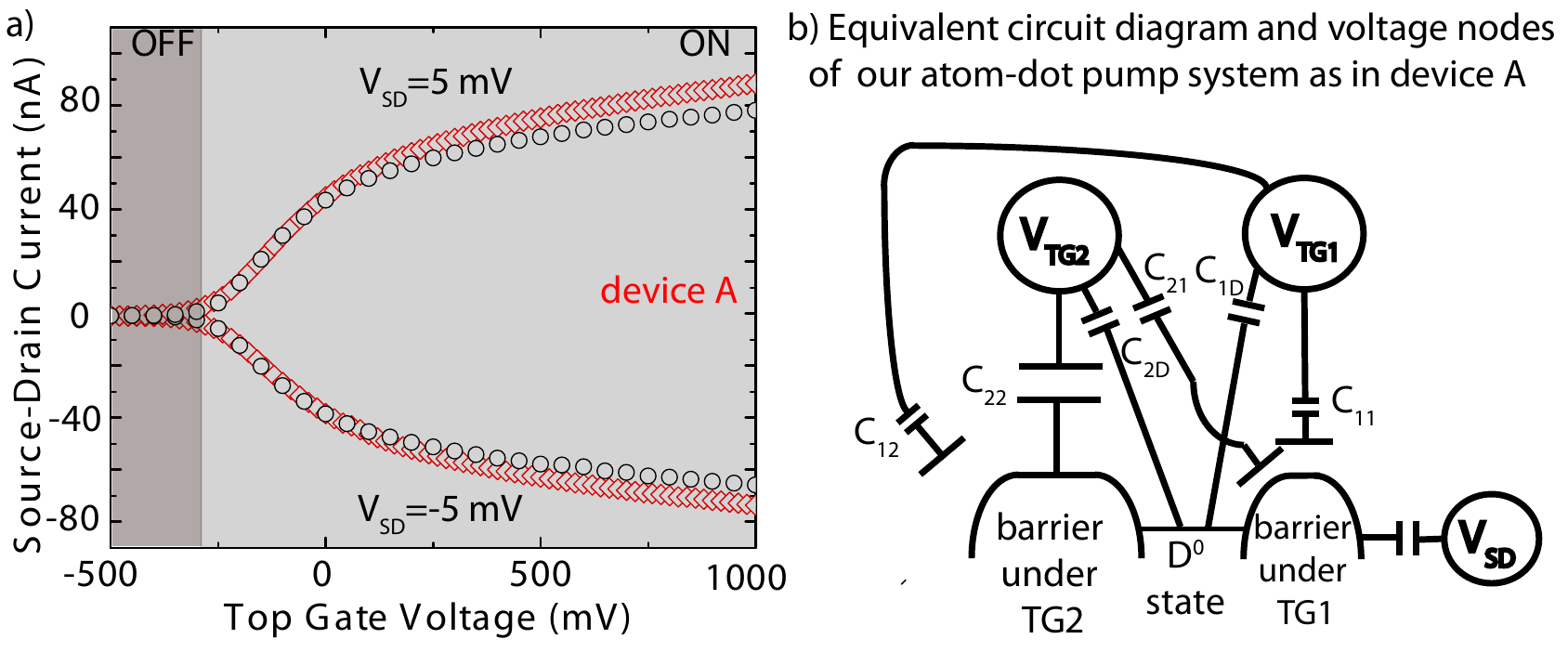}
\caption{a) Room Temperature source-drain current versus TG1 (TG2) traces (with TG2 (TG1) grounded), with circles and diamonds respectively and $V_{SD}$=$\pm$ 5 mV for Device A. As expected for the dopant-atom signature \cite{Wac193},  both the TG's have $V_{th}$~$\lesssim$~0~mV. Furthermore both TG's have a almost symmetric control on the channel and can, therefore, one independently from the other, switch the transistor to the ON-state. Similar data are obtained for all the devices studied in this work. b) Equivalent circuit diagram and voltage nodes of our atom-dot pump system as in device A. $C_{11}$ ($C_{22}$) is the coupling between the barrier under TG1 (the barrier under TG2) and TG1 (TG2), $C_{1D}$ ($C_{2D}$) is the coupling between the TG1 (TG2) and the state and $C_{12}$ ($C_{21}$) is the cross-coupling between TG1 (TG2) and the barrier under TG2 (the barrier under TG1), respectively.}
\label{fg:SUPPPUMPING1}
\end{center}
\end{figure*}

\begin{figure*}
\begin{center}
\includegraphics[width=172mm]{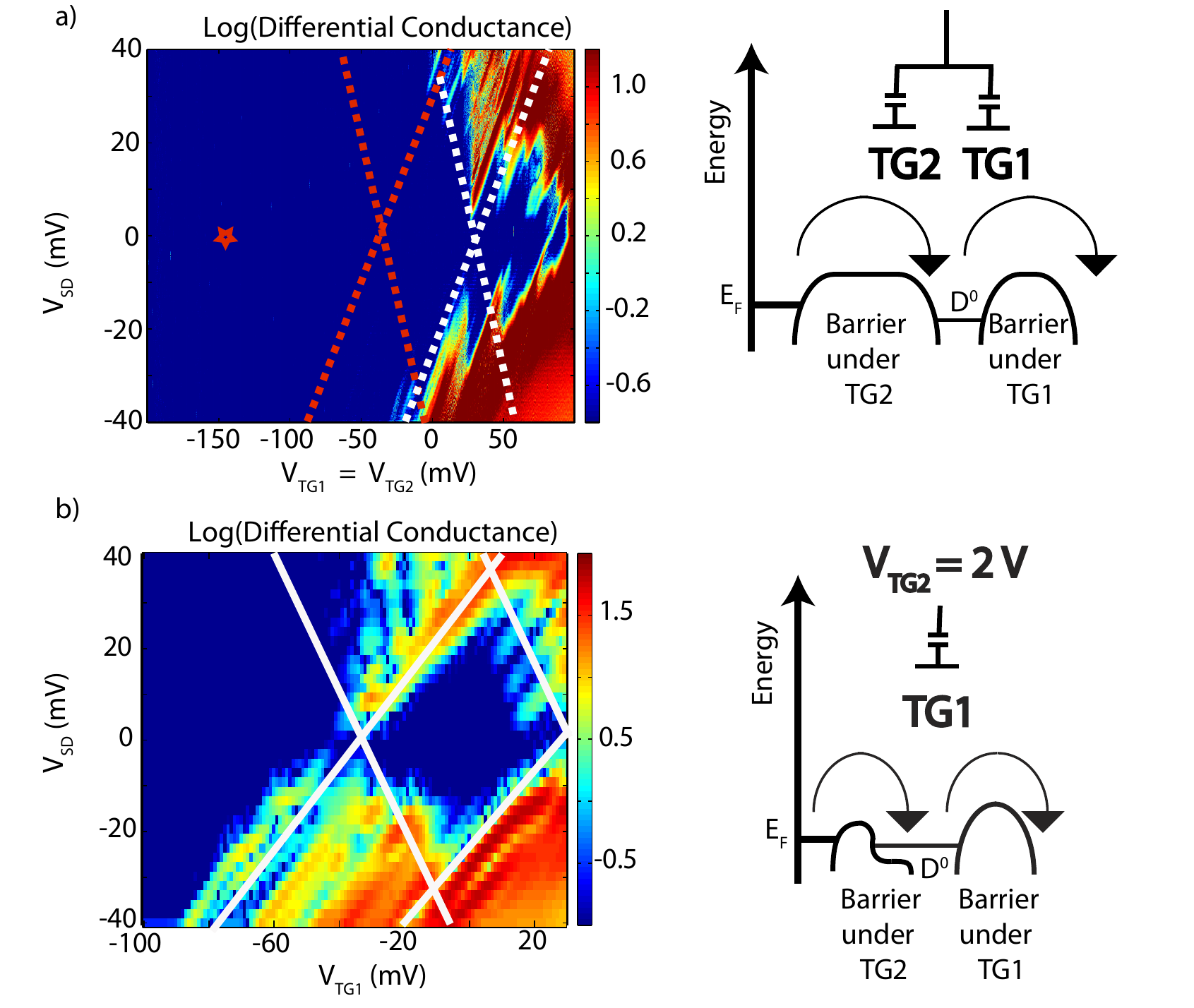}
\caption{a) Coulomb Blockade (CB) stability diagram, acquired at 4.2~K, of the differential conductance, $G$=$dI_{SD}$/$dV_{SD}$, versus the voltage of the two top gates addressed together and $V_{SD}$. Due to an estimated $E_{C}$ ranging between 30~meV and 40~meV, the first peak is schematically indicated by means of the two red dotted lines, 80$-$90 mV from the second one (i.e. indicated by white dotted lines) and can be associated to the $D^{0}$ charge state \cite{Sel206805} of a single $P$ atom with an isolated ground state \cite{Roc206812}. b) 2D Stability diagram of the differential conductance of a region of gate voltage schematically indicated with the red dotted lines in a), but this time with $V_{TG2}$~=~2~V. From this figure a charging energy ($E_{C}$) of around 35 meV can be estimated, strongly pointing towards the atomic nature of the state we are dealing with \cite{Sel206805}. As shown in the schematics at right of the figure, initially (top schematic) the barriers are not transparent and asymmetric, probably due to the non perfect central position of the $P$ atom if compared to the two gates. This poor barriers' transparency situation is optimal for the weak coupling regime necessary to the RF pumping but does not allows the observation of the full Coulomb diamond signature in DC transport. However, when one of the two top gate is keep fixed at a positive value (bottom schematic), the transparency of both the barrier is sufficiently increased and the diamond signature can be observed.}
\label{fg:PUMPING2}
\end{center}
\end{figure*}

The devices used in our experiments are fabricated on a Complementary-Metal-Oxide-Semiconductor (CMOS) platform \cite{Roc206812,Roc032107,Wac193} making them fully compatible with one of the most successful technologies of modern times. An adapted \cite{Roc032107} fully depleted-silicon on insulator (FD-SOI) technology (see Fig.~\ref{fg:PUMPING1}) is used to fabricate n-metal-oxide-semiconductor~field-effect-transistors (n-MOSFET's). The 20 nm thick SOI channel is initially $P$ doped with a background concentration of $10^{18}$ $cm^{-3}$ and then is etched to form the channel of the device which is sitting on top of a 150 nm thick buried oxide (BOX), as shown in Fig.~\ref{fg:PUMPING1}b). The device fabrication is then completed by using a technology similar to that used in commercial trigate SOI MOSFETs \cite{Roc032107}. The only difference is that, in our devices, not only one gate (polycrystalline silicon) is wrapped around three sides of the channel (see Fig.~\ref{fg:PUMPING1}b)), but two are positioned in series (see Fig.~\ref{fg:PUMPING1}a)) along the direction of electron transport. Finally, to protect the channel from the high doping doses (HDDs) necessary for the formation of the source and the drain regions, 15 nm thick $Si_3N_4$ spacers (see Fig.~\ref{fg:PUMPING1}a)) are formed around the two gates~\cite{Roc206812} and the HDDs implantation is performed at an angle of 55 degrees. This configuration prohibits $As$ dopants from the HDD from reaching the portion of the channel between the two gates. Furthermore, within this channel, the original $P$ doping yields only a few active $P$ atoms, which provide the $D_{0}$ states shown in Fig.~\ref{fg:PUMPING2}, hence allowing the observation of the SAT behaviour \cite{Sel206805,Roc206812,Mor345,Fue242,Pla541,Tet046803}. Note that in our devices the $P$ atoms are present in the centre of the channel only due to the initial background doping procedure discussed above. Furthermore, as shown in Fig.~\ref{fg:SUPPPUMPING1}a) and Fig.~\ref{fg:PUMPING2}, room and low temperature \cite{Wac193,Roc206812} electrical characterisation can be used to study the presence of this discrete number of isolated dopants in the channel and their effects on the transport characteristics.

The low temperature measurements were performed by mounting the device on a stage that is dipped directly in liquid helium and thus at 4.2~K. A low noise battery operated measurement setup (see Ref.~\cite{Van4394,Vin123512,Koppens}) was used to measure the source/drain current and to apply the voltages. As the amplifier is at room temperature, the absolute noise floor of the IV-converter is limited by the Johnson noise of the feedback resistor, which is given by $\sqrt{\frac{4k_{B}T}{R_{f}}}$~=~4~$\frac{fA}{\sqrt{Hz}}$. To apply the sinusoidal RF input to one gate an Agilent 8648C source was used. An important aspect of our setup is that a cryogenic bias-T \cite{Gib1}, in red in the circuit schematic of Fig.~\ref{fg:PUMPING1}a) and \ref{fg:PUMPING1}b), is used to add this RF voltage to one of the two DC voltages applied to both the gates; this in a \textit{single-parameter charge pumping configuration} \cite{Gib1,Fuj042102,Kae012106}.

\begin{figure*}
\begin{center}
\includegraphics[width=172 mm]{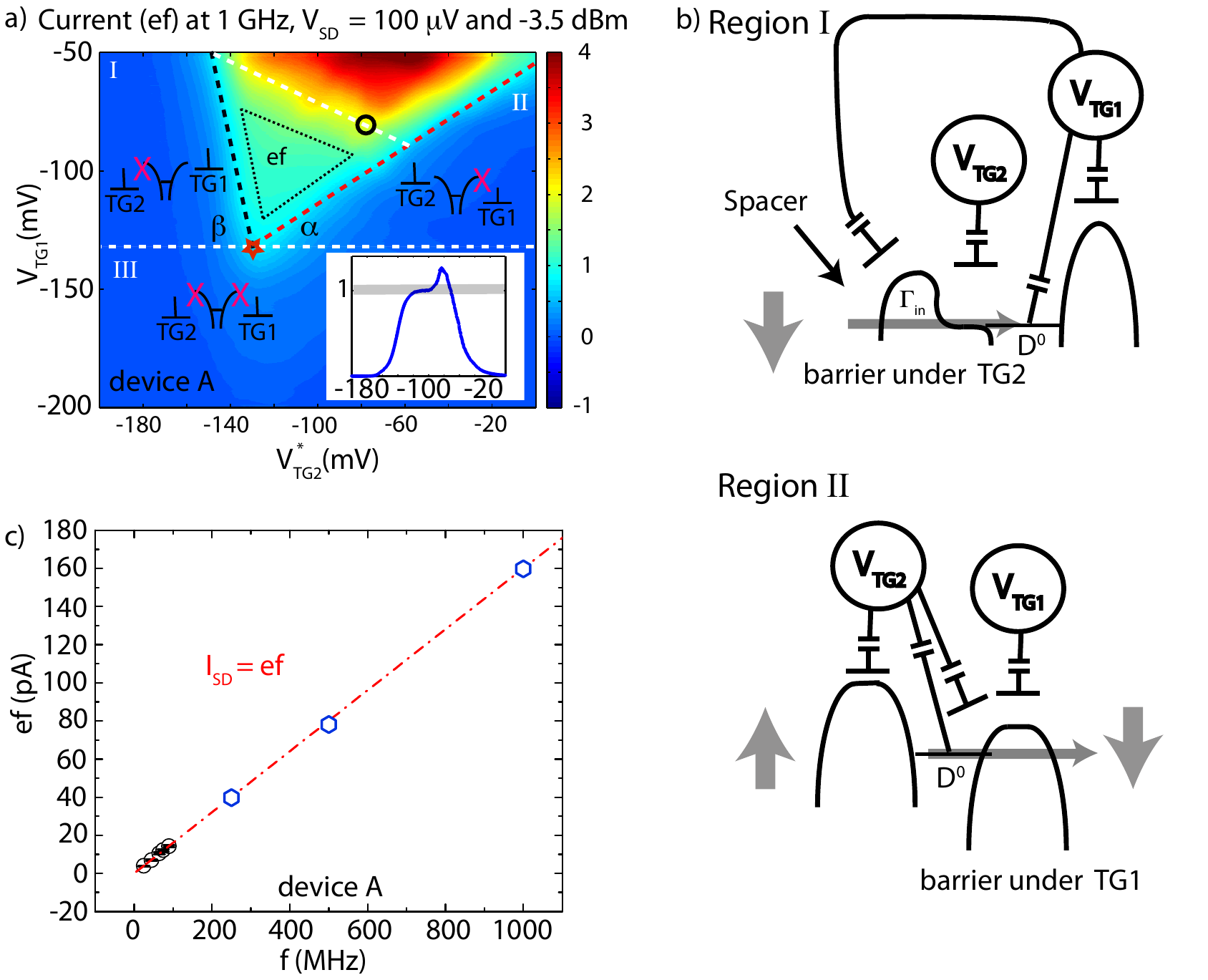}
\caption{ a) 2D stability of the $I_{SD}$ versus $V_{TG1}$ and  $V_{TG2}$ taken at 4.2~K and by adding a sinusoidal $V_{RF}$ to TG2 of device $A$. The power used at 1~GHz is -3.5~dBm, which, for our lines is equivalent to having a $V_{RF}$~$\approx$~0.1 V. The distance between the red star (indicating the onset of the pumping current) and the crossing of the red dotted lines (indicating the $D^{0}$ state of the $P$ atom) in Fig.~\ref{fg:PUMPING2}a) is in agreement with the used $V_{RF}$. Furthermore, the distance between the red star and the black circle related to the second electron state is around 80 mV which, again, is in agreement with the distance between the crossing of the red dotted lines and the one of the white dotted lines (indicating a second electron state) observed in Fig.~\ref{fg:PUMPING2}a). These facts demonstrate that the transport signature we observe is  related to charge pumping through the same electron state discussed in Fig.~\ref{fg:PUMPING2}. Furthermore, as explained in the text, the angle $\beta$, defining region \textrm{I}, gives an indication of the coupling between TG1 and the state. The angle $\alpha$, defining region \textrm{II}, gives an indication of the coupling between TG2 and the state. Furthermore, $\alpha$ and $\beta$ give also an indication of the control that each gate has on the barrier located under the opposite one, respectively. The white inset illustrates the correct reaching of the $<$n$>$~=~1 step for a fixed $V_{TG1}$~=-~90~mV. b) Schematic of the tunneling processes at the limit between region \textrm{I} and the pumping region (black dashed line in a)) and at the limit between the pumping region and region \textrm{II} (red dashed line in a)). An equivalent circuit diagram indicating all the couplings and all cross couplings present in device $A$ is also shown in Fig.~\ref{fg:SUPPPUMPING1}b). c) Figure showing how device $A$ follows the ideal $I_{SD}$~=~$ef$ behaviour up to 1~GHz. Different colours in the points indicate different conditions of acquisition of the data. The details on each point in this curve are discussed in Fig.~\ref{fg:SUPPPUMPING2} of section~\ref{sec:A3}. The red line is just a guide of eye on how the points align compared to the ideal behaviour.}
\label{fg:PUMPING3}
\end{center}
\end{figure*}

Our double gate (TG1 and TG2 in Fig.~\ref{fg:PUMPING1}a)) configuration allows the control of both the left and the right barriers, thereby confining the electrons into the atomic potential using an approach similar to the one recently implemented elsewhere~\cite{Gib1,Blu343,Fuj042102,Kae012106}. However, differently from previous ones~\cite{Gib1,Blu343,Kae012106}, in our approach each gate is capacitively coupled with the dopant state (in a way similar to that of the global top gate used in ref.~\cite{Fuj042102} or of the back gate in ref.~\cite{Sie3841}) and, to some extent, can also cross-couple to the barrier under the opposite gate, e.g. see schematic in Fig.~\ref{fg:SUPPPUMPING1}b) and Fig.~\ref{fg:PUMPING3}b). This makes it possible to observe a pumping behaviour not previously observed in conventional pumps. As an example, in our devices the distinction between "entrance" and "exit" gates is not as sharp as in conventional QEPs \cite{Gib1,Kae012106}, which explains why both the gates can be used for the evaluation of the quantisation (pumping) effects.

Two QEPs, i.e. devices $A$ and $B$, structurally similar to the one shown in Fig.~\ref{fg:PUMPING1}a) and Fig.~\ref{fg:PUMPING1}b) and having channel widths ($W$s) equal 60~nm and 800~nm and gates lengths ($L_{g}$s) equal 50~nm and 20~nm, respectively, have been selected between the many available and studied at 4.2 K.

Initially, many devices are measured at room temperature. Source-drain current versus TG1 (TG2) traces are acquired by keeping the other gate, TG2 (TG1), grounded as shown in Fig.~\ref{fg:SUPPPUMPING1}. The following signatures (e.g. see Fig.~\ref{fg:SUPPPUMPING1}) are used to select the good devices amongst the large pool of fabricated ones \cite{Roc206812,Wac193}: 

\begin{enumerate}
  \item Negative threshold voltages, $V_{th}$ $\lesssim$ 0 mV, indicating the dominance by a discrete number of dopant atoms in the sub-threshold region of transport \cite{Wac193}.
  \item Almost symmetric behaviour of the gates (indicating a good control of the channel by both the gates) and good results in the lithography of the poly-gates themselves. 
  \item The capability that each gate has, independently from the other, to control the channel (and to turn the transistor to the ON-state). This, at low temperature, results in the important feature that each gate can, to some extend, have some cross-coupling with the barrier under the opposite gate (as schematically shown in Fig.~\ref{fg:SUPPPUMPING1}b)). 
\end{enumerate}

The data in Fig.~\ref{fg:SUPPPUMPING1}a) are related to the room temperature characterisation of device $A$ studied in the main part of the manuscript, a similar selection process has been followed for device $B$. The data in Fig.~\ref{fg:PUMPING2} are related to 4.2~K characterisation of device $A$. For devices having the distance between the two top gates ($S_{gg}$)~=~50~nm, such as device $A$ and device $B$, only for $V_{TG1}$ and $V_{TG2}$~$\gtrsim$~100 mV can \textit{artificial atom} systems (i.e. QDs) be induced under the two top gates. Therefore, the sub-threshold (for $V_{TG1}$ and $V_{TG2}$~$\lesssim$~100 mV) current signature can only be dopant related, see also similar results in Ref.~\cite{Roc206812,Wac193,Lan763,Roc1581}. Fig.~\ref{fg:PUMPING2} indicates that our approach enables the few electrons regime~\cite{Roc206812} to be accessed and gives a first indication of the fact that, for device $A$, the first charge state ($D^{0}$) of a single isolated $P$ dopant atom (located in the central region of the channel) is addressed. As our pumping approach requires low transparency of the tunnel barriers, the exact quantification of the $E_{C}$ in the CB diamonds of Fig.~\ref{fg:PUMPING2}a) is not straightforward. The estimation of $E_{C}$~$\approx$~35~meV obtained in this Fig.~\ref{fg:PUMPING2}a), however, was confirmed  in the strong coupling regime \cite{Tet046803} measurements of Fig.~\ref{fg:PUMPING2}b). This strong coupling regime is not optimal for pumping but it allows the observation of the CB diamond for a gate position similar to the one where it is only partially visible in Fig.~\ref{fg:PUMPING2}a).

These results agree with the analysis of the AC data discussed in section~\ref{sec:S3} and with recent observations by other groups~\cite{Roc206812,Wac193,Lan763,Roc1581}. Furthermore, for our pumping experiments, the channel thickness ($T$) and $S_{gg}$ have always been fixed at 20 nm and 50 nm respectively because, unlike $W$ and $L_{g}$, the parameter $S_{gg}$ has proven to be important. As an example, the absence of transport in the sub-threshold region and a regular double quantum dot (DQD) signature~\cite{Wie1} in the region of inversion, i.e. for $V_{TG}$'s~$\gtrsim$~100 mV, have been observed for other measured devices having $S_{gg}$~$>$~50~nm (see section~\ref{sec:A2}). These facts are an indication of the absence of transport through isolated dopants and indeed pumping through isolated dopants has proved not to be possible in this latter case.


\section{$\textbf{Radio frequency (RF) measurements}$}
\label{sec:S3}

\begin{figure*}
\begin{center}
\includegraphics[width=172mm]{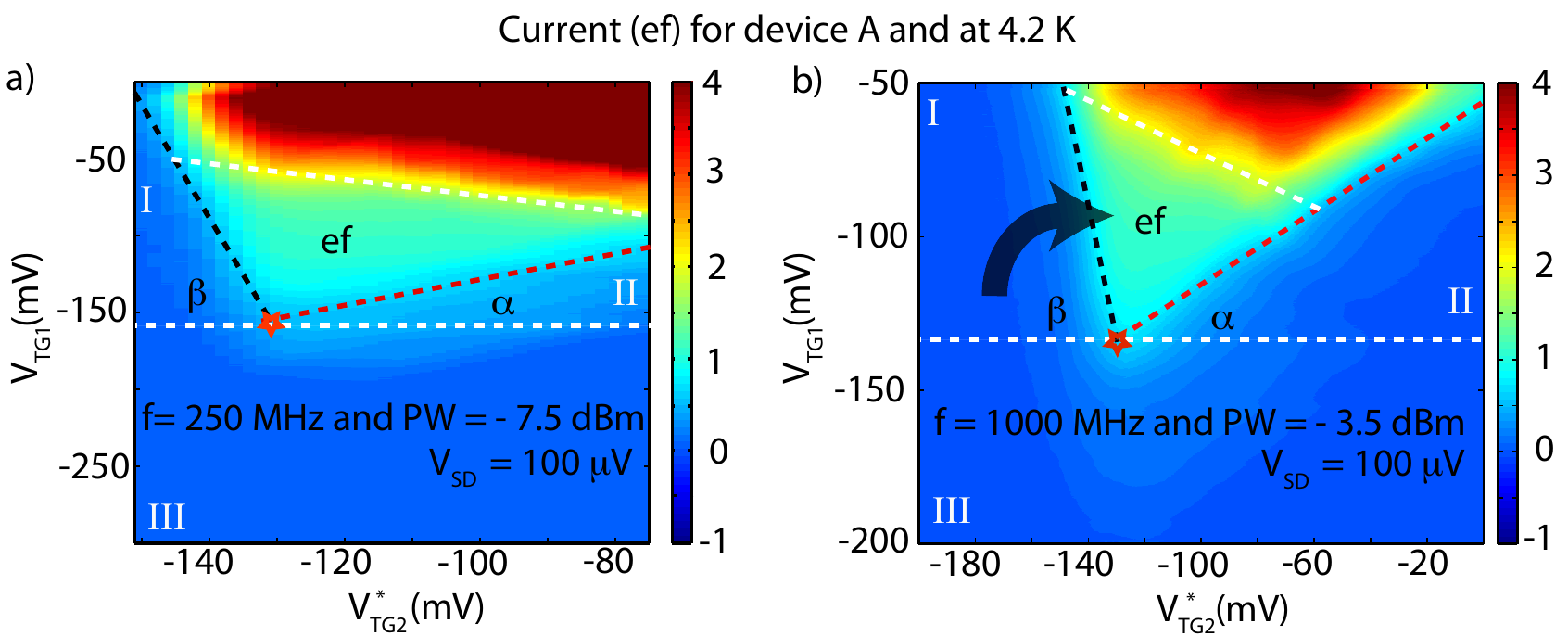}
\caption{Stability of the pumping current at 4.2 K for a) $f$~=~250~MHz and PW= -7.5 dBm. b) $f$~=~1~GHz and PW= -3.5 dBm. The symbol are as in Fig.~\ref{fg:PUMPING3}a). From these data it is possible to extract that $\tan\beta$ is $\cong$~6.8 and 8.5 for $f$~=~0.25~GHz and 1~GHz, respectively. This observation confirm the trend of increasing $\beta$ following the increase of $f$ as discussed in the text.}
\label{fg:SUPPPUMPING4}
\end{center}
\end{figure*}

\begin{figure*}
\begin{center}
\includegraphics[width=172mm]{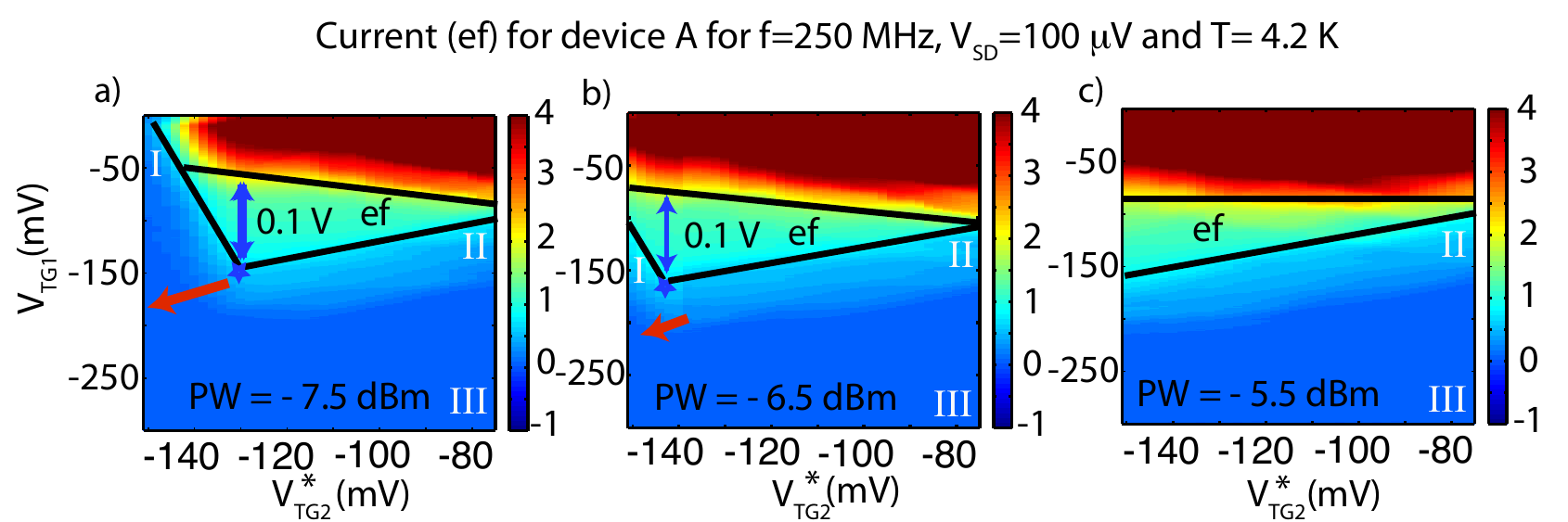}
\caption{Stability of the pumping current at $f$~=~250~MHz at 4.2 K and for a) PW= -7.5 dBm, b) PW= -6.5 dBm and c) PW= -5.5 dBm, respectively.}
\label{fg:SUPPPUMPING3}
\end{center}
\end{figure*}

\begin{figure*}
\begin{center}
\includegraphics[width=172mm]{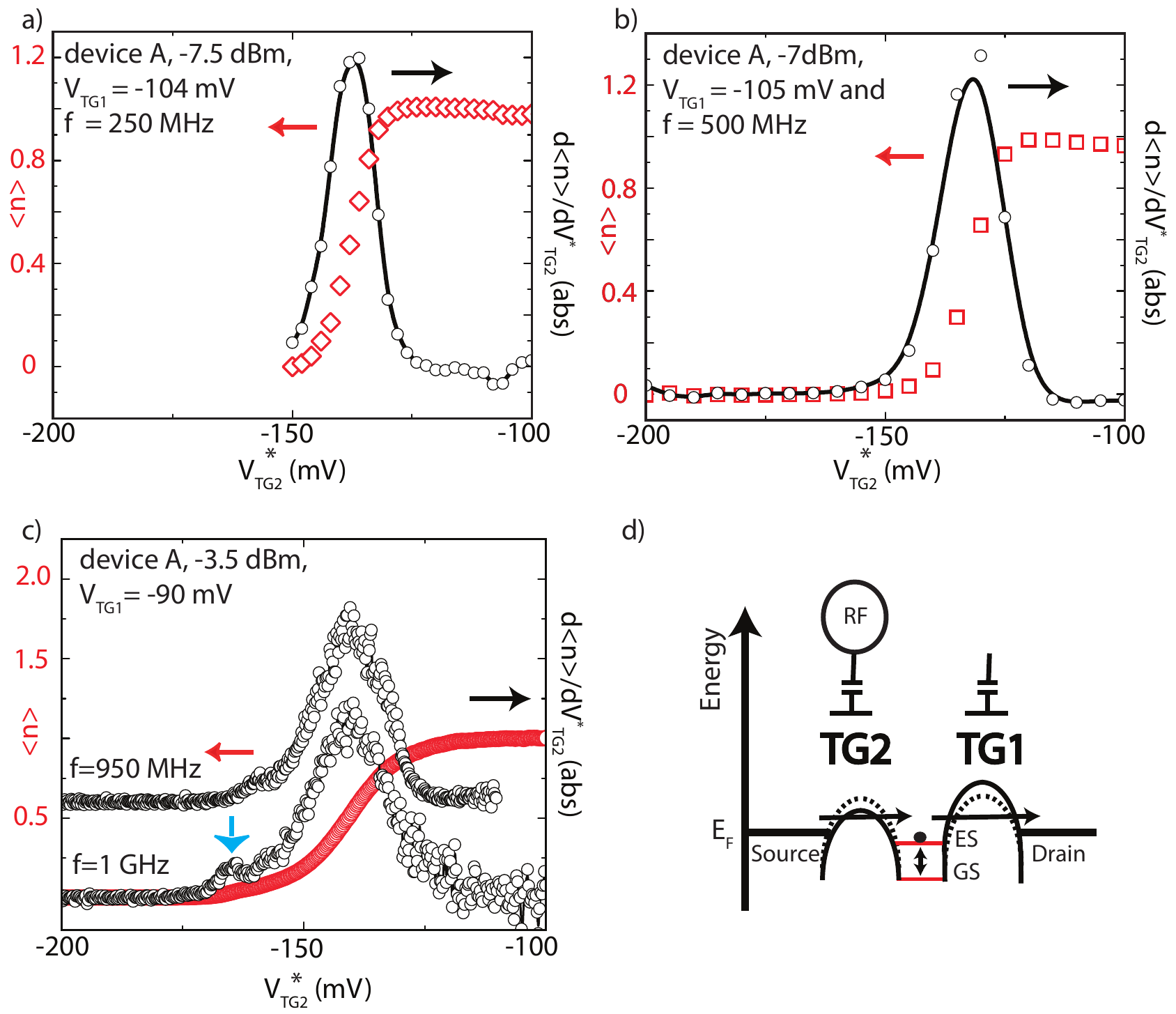}
\caption{The current data of device $A$ for $f$~=~250~MHz, ~500~MHz and 1~GHz are shown in red in a), b) and c), respectively. For each of these curves the reaching of $<$n$>$~=~1 is shown. The $\frac{d<n>}{dV^{*}_{TG2}}$  data are shown for each frequency in black. For a) and b) the curves do not show the effects related to non-adiabatic excitations \cite{Kat126801}, but, as outlined by the blue vertical arrows, these start to play a role in the GHz range of c). From the $\frac{d<n>}{dV^{*}_{TG2}}$ curve for  $f$~=~950~MHz, also shown in section c) of the figure with its x and y-axis translated for clarity, it is possible to observe that the small peak rapidly disappear when $f$ decreases and becomes totally absent in the data of section b) taken at $f$~=~500 MHz. Hence this confirm it is related to non-adiabatic excitations \cite{Kat126801}. The onset of non-adiabaticity may also be related to the asymmetry effects shown in Fig.~\ref{fg:PUMPING5}a). c) d) Schematic describing how the non-adiabatic excitation effects could start to affect our system at 1~GHz.}
\label{fg:PUMPING4}
\end{center}
\end{figure*}

\begin{figure*}
\begin{center}
\includegraphics[width=172mm]{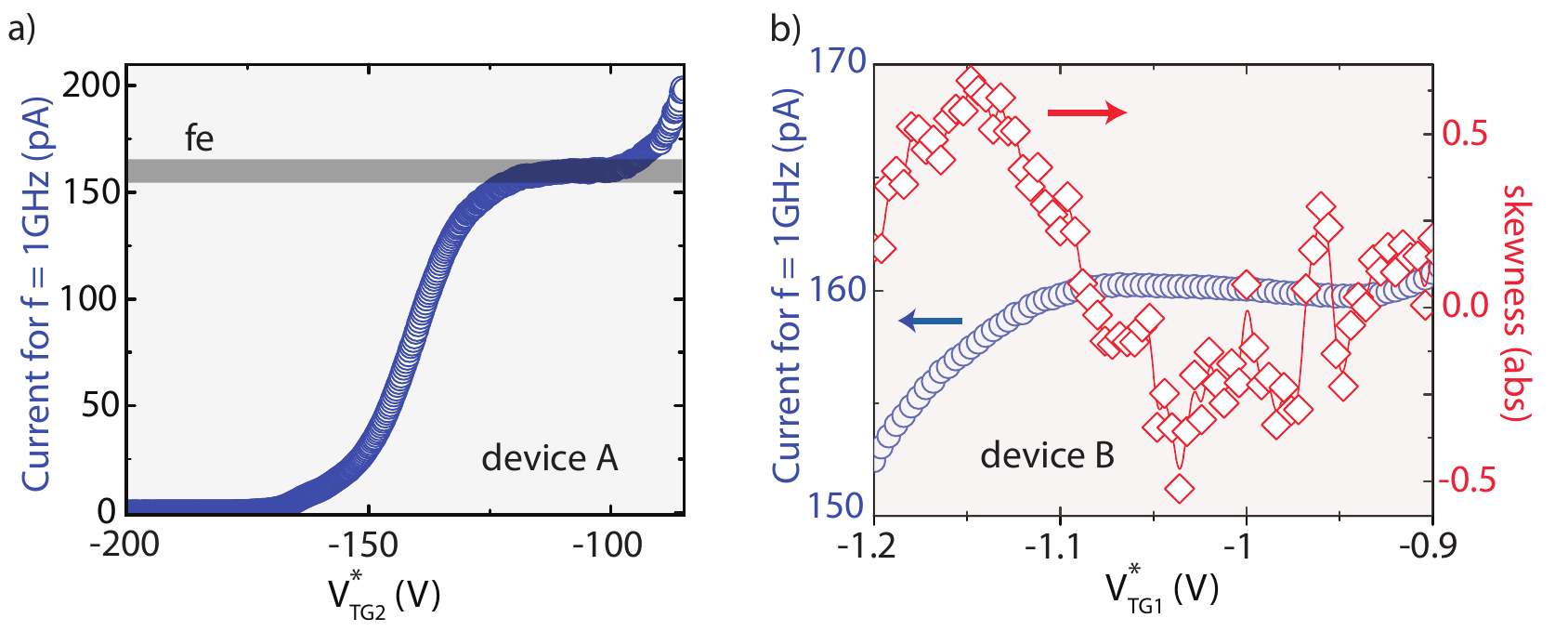}
\caption{ a) Pumping current generated with device $A$ acquired at 4.2~K and for $f$~=~1~GHz (blue circles) as in Fig.~\ref{fg:PUMPING4}c). b) Pumping current generated with device $B$ acquired at 4.2~K and for $f$~=~1~GHz (blue circles) and curve of the values of the asymmetry of the probability distribution for the same data (red diamonds). The skewness for a sample of $n$ values is defined by $\frac{\frac{1}{n}~\sum~(x_i-\overline{x})^3}{(\frac{1}{n}~\sum~(x_i-\overline{x})^2)^\frac{3}{2}}$, with $\overline{x}$ being the mean value. As described in the text, the data have been acquired by applying a fixed 5.715~V to the back gate (BG). By looking at the data of the skewness, it is possible to observe that, as for device $A$, also for this device $B$ the asymmetry of the data \cite{Fuj042102} at the onset of the pumping current, appears at the GHz frequencies.}
\label{fg:PUMPING5}
\end{center}
\end{figure*}

We can now turn to the radio frequency (RF) measurements of device $A$. In Fig.~\ref{fg:PUMPING3},  Fig.~\ref{fg:SUPPPUMPING4},  Fig.~\ref{fg:SUPPPUMPING3} and Fig.~\ref{fg:PUMPING4}, the data related to the generation of pumping currents when a sinusoidal voltage excitation, $V_{RF}$, is provided to TG2 are shown. The asterisk, "$\ast$", used for the AC data, indicates for each pump which of two gates is excited by RF oscillations. Taking advantage of the knowledge gained in the DC transport experiments illustrated in Fig.~\ref{fg:PUMPING2}, the top gates are tuned to values far from the ones of the DC transport regime, i.e. in the region indicated with the red star in Fig.~\ref{fg:PUMPING2}a). For these voltage values, transport can only be related to the transiting of electrons through the atom-dot following the pumping scheme introduced in Fig.~\ref{fg:PUMPING1}c). This is also illustrated in Fig.~\ref{fg:PUMPING3}c), where the reaching of unity steps proportional to the frequencies of excitation, $f$, ranging from~25~MHz to~1~GHz and the subsequent following of the expected $I_{SD}$=$fe$ law are shown. Furthermore, in our pumps, the sign of the current does not depend on the sign of $V_{SD}$ but on which gate is selected to be RF excited \cite{Fuj042102,Fle155311}, i.e. the data of Fig. \ref{fg:PUMPING5} related to device $B$ have been acquired by using $V_{SD}$~=~-~2~mV. The independence of the current from the source/drain polarity is another strong indication of the pumping nature of a current \cite{Fuj042102}. These last facts allow us to illustrate some of the great advantages of our SAT based single-parameter configuration, which, by combining an isolated ground state with large $E_{C}$'s, allows the extension of the non-adiabatic pumping regime \cite{Kae012106,Fuj042102,Kae153301} for at least three orders of magnitude, i.e. from the MHz to the GHz regime of operations. This also occurs since a careful control of the phase of the excitation is not necessary \cite{Fuj042102}.

As shown below, the analysis of the triangular shape observed in Fig.~\ref{fg:PUMPING3}a) (and in Fig.~\ref{fg:SUPPPUMPING4} and Fig.~\ref{fg:SUPPPUMPING3}) brings further evidence that the pumping signature studied for device $A$ is dopant related. In region \textrm{III}, the AC transport is never possible. In region \textrm{I}, opposite to angle $\beta$, the current is limited because the electrons are blocked at the entrance of the pump (stage 1 in Fig.~\ref{fg:PUMPING1}c)). This is true unless TG1 can compensate for the lack of action of TG2 in addressing the barrier under it and the state~\cite{Lei911}. This also justifies the association between $\beta$ and the value of coupling between TG1 and the state and the value of the cross-coupling between TG1 and the barrier under TG2, see Fig.~\ref{fg:PUMPING3}b) for a schematic. The estimated value for $\beta$~(the voltage ratio $\frac{\bigtriangleup V_{TG1}}{\bigtriangleup V_{TG2}}|_{\textrm{I}}$~=~$\tan\beta$ is~$\approx$~8.5, hence $\beta$~$\cong$~83.3~degrees) indicates that the black dotted onset line is 6.7~degrees more tilted in the anti-clockwork direction compared to the case of having the state coupled only to TG2 and no cross-coupling between TG1 and the barrier under TG2, i.e. this case should lead to $\beta$~=~90~degrees~\cite{Lei911}. To clarify this aspect we can see if the observed dependence of $\beta$ to $f$ agrees with it. It should be noted that expected $\beta(f)$ can be extrapolated, and so as to observe full quantisation, the entrance rate ($\Gamma_{in}$ in Fig.~\ref{fg:PUMPING3}b)) in the first half of the cycle must be large compared to $f$~\cite{Lei911}. So if the value of $f$ increases, $\beta$ is expected to increase, i.e.: it becomes more difficult for TG1 to be effective in correcting the lack of action of TG2 in lowering the barrier. This occurrence is demonstrated in Fig. \ref{fg:SUPPPUMPING4}, where it is demonstrate that the angle $\beta$ increases following an increase of $f$. Indeed, being $\tan[\beta(f=250~MHz)]$~=~6.5 and $\tan[\beta(f=1~GHz)]$~=~8.5, we have $\beta(f=250~MHz)$~=~81.2~degrees and $\beta(f=1~GHz)$~=~83.3~degrees.

In region \textrm{II}, opposite to angle $\alpha$, the pumping current is limited when the electrons are blocked at the exit of the pump (stage 3 in Fig.~\ref{fg:PUMPING1}c)). This happens because during each cycle the barrier under TG2 is so low that TG1 needs to be larger than previously to let the electrons leave the atom to the drain, see Fig.~\ref{fg:PUMPING3}b). When the DC value of $V_{TG2}$ decreases, the exit of the electrons and the complete cycle are facilitated and becomes possible for lower values of $V_{TG1}$. This justifies the association between $\alpha$ and the value of the coupling present between TG2 and the state and the value of the cross-coupling between TG2 and the barrier under TG1. The value for $\alpha$~(the voltage ratio~$\frac{\bigtriangleup V_{TG2}}{\bigtriangleup V_{TG1}}|_{\textrm{II}}$~=~$\tan(90-\alpha)$ is~$\approx$~1.69, hence $\alpha$~is~$\cong$~31~degrees) is non-zero. $\alpha$ and $\beta$ can only provide an insight into the complicated coupling and cross-coupling effects present in our system (see Fig.~\ref{fg:SUPPPUMPING1}b)), however, there is an important information that can be extracted from these: the two voltage ratios associated with $\alpha$ and $\beta$ are of the same order of magnitude and hence the position of the state can be inferred somewhere at the centre of the channel.

Lastly, the strong frequency dependence of the angles $\alpha$ and $\beta$ observed in Fig.~\ref{fg:SUPPPUMPING4} could be also related to the capacitive coupling between the gates, as schematically shown in Fig.~\ref{fg:SUPPPUMPING1}b). This because, during the RF measurements, the "fixed" gate $V_{TG1}$ will oscillate due to stray coupling with the driven gate $V_{TG2}$ and the amplitude and the phase of this parasitic oscillation will be frequency dependent.

By following the black and the red dashed lines towards more negative values of $V_{TG1}$ it is possible to observe that they do cross (the red star in Fig.~\ref{fg:PUMPING3}a) and in Fig.~\ref{fg:SUPPPUMPING4}). In conventional electron pumps \cite{Gib1,Kae012106,Fle155311,Lei911}, the confinement site is created by applying negatively voltages above a two-dimensional electron gas (2-DEG) and the energy state of the QD is completely manipulated by the barriers and crosses the Fermi level when the barriers are already transparent enough to observe pumping current \cite{Lei911}. Moreover, the state used for pumping is localised in a region that comprises many atomic sites and as a consequence, for these 2-DEG induced QDs~\cite{Gib1,Kae012106,Lei911}, there is always a certain range of  $V_{TG1}$ and  $V_{TG2}$ where the onset of the pumping transport can be observed (i.e. the shape of the plateau in the 2D stability is trapezoidal and not triangular). This observation reflects the fact that for these confining potentials the states are not strongly localised and can tolerate some deformation from the gates before transiting from one regime to another. However, when the localisation of the state is increased, for example following the application of a magnetic field perpendicular to the plane~\cite{Gib1,Kae012106,Fle155311}, a drastic reduction of this range of tolerance, is observed. Indeed, as shown in Fig. 2 of ref~\cite{Kae012106}, the trapezoidal shape of the 2D stability approaches the triangular one with the application of a magnetic field.

The fact that in our system this region of tolerance is reduced to a small region of the 2D stability (red star in in Fig.~\ref{fg:PUMPING3}a) and in Fig.~\ref{fg:SUPPPUMPING4}), even without the presence of a magnetic field, indicates that the potential confining the electrons is localised in a few atomic sites, i.e. is related to an isolated donor atom \cite{Sel206805} and not a QD. This occurs because, for an atom pump, the barriers are not transparent enough when the state crosses the Fermi level (as also seen in Fig.~\ref{fg:PUMPING2}a)) and the crossing line is only seen for the second electron (outlined by the black circle in Fig.~\ref{fg:PUMPING3}a)). Furthermore, as shown in the Fig.~\ref{fg:SUPPPUMPING3} for $f$~=~250~MHz, following an increase of the RF power (PW) from~-7.5 dBm to~-5.5 dBm, the triangular shape equivalent to the one observed in~\ref{fg:PUMPING3}a) does not seems to be dramatically deformed but just shifted towards more negative gate voltage values. This last observation is dramatically different from the ones attributed to conventional QEPs, see ref~\cite{Gib1,Kae012106,Lei911} and references therein. This is again in agreement with the idea that a strong local potential, having a shape less affected by the gates if compared to conventional QEPs, is responsible for the confinement of our pumping electrons, making that the increase of PW can move the onset of the pumping current to more negative of the DC gate voltages. Lastly, the second plateau observed in Fig.~\ref{fg:PUMPING3}a) (black circle) can be associated with the $2e$ state ($D^{-}$ state \cite{Lan136602}) of the first atom or with the $1e$ state of a second atom. For this reason, in this paper, we only study the first quantisation step.

To conclude the discussion on device $A$, Fig.~\ref{fg:PUMPING4} shows the data for the RF currents and the derivative of the currents for different frequencies. From the data of Fig.~\ref{fg:PUMPING4} it is possible to extrapolate that, in our system, the onset of the pumping current starts to be mildly affected by non-adiabatic excitations \cite{Kat126801,Fle155311} only for $f_{SAT}$~$\approx$~1~GHz and, as also shown in Fig.~\ref{fg:PUMPING3}, little degradation of the quantisation is observed even for this range of frequencies. If we compare this $f_{SAT}$ to the onset of non-adiabatic excitations observed in QD-QEPs \cite{Kat126801,Fle155311}, i.e. $f_{QD}$~$\approx$~200~MHz, we have another indication of the fact that the system addressed in our experiments is characterised by higher energies if compared to conventional QEPs. Indeed, for a SAT \cite{Roc206812}, all the energy scales (e.g. $E_{C}$~$\approx$~30-40~meV and $\triangle E_{Atom}$~=~$E_1$-$E_{GS}$~$\approx$~10~meV, with $E_{GS}$ being the ground state and $E_1$ the first excited state) are higher if compared to the ones of a QD-QEP \cite{Kat126801,Fle155311} and it is therefore possible to observe charge pump quantisation at 1 GHz. Plus, even if our SAT pumps are operating at relatively high temperatures ($T_{op}$~$=$~4.2~K), as the condition $\exp(-\frac{E_{C}}{k_{B}T_{op}})$~$\sim$~0 is still holding, thermal errors \cite{Zim5254} can be ignored. In conclusion, in our SAT system, high $E_{C}$ (i.e.~immunity to thermal effects) combined with the isolated ground state (i.e.~immunity to non-adiabatic excitations) allows the observation of the quantised non-adiabatic regime~\cite{Kae153301} for a wide range of frequencies. This is telling us that the use of SAT systems~\cite{Sel206805,Roc206812,Fue242} having larger~$E_{C}$ and larger~$\triangle E_{Atom}$ signatures, if compared to the ones associated to the present system, could allow further improvements of the pumping performances.

From the data of device $B$, selected and characterised via room and low temperature DC measurements identical to device $A$, we gain a more precise idea of the robustness of our SAT approach. As the setup used in our experiments is a state-of-the-art custom build ultra low noise apparatus \cite{Van4394,Vin123512,Koppens}; it is able to provide a noise floor of $\frac{4~fA}{\sqrt{Hz}}$ when a 1~G$\Omega$ feedback resistor is used in the amplification and $\frac{12~fA}{\sqrt{Hz}}$ when a 0.1~G$\Omega$ feedback resistor is used (as for all our data). Hence, in this configuration, a 10 Hz bandwidth will lead to the association of 40~$fA$ of error to each point. Furthermore, the current data of this Fig.~\ref{fg:PUMPING5}b) has been obtained by repeating the same 10 Hz bandwidth trace $\approx$~1200 times in 24~$h$ and therefore the effective bandwidth could be as low as $\frac{10~Hz}{\sqrt{1200}}$. However, the optimal bandwidth was estimated by also taking into account the long time instabilities associated with our room temperature acquisition system. In the case of the current data of Fig.~\ref{fg:PUMPING5} b), the optimal trade-off leads to a bandwidth around 1.6 Hz and to an estimated error of around 15 fA for each point in the plateau. Indeed, a standard error of the mean of this order of magnitude has been also extracted from the data in the plateau region of this Fig.~\ref{fg:PUMPING5}b). These facts give an indication on the robustness and on the reproducibility that can be associated to our system. Furthermore, the data of Fig.~\ref{fg:PUMPING5} are important as they demonstrated that we can obtain the $I_{SD}$~=~$fe$ quantisation for $f$~=~1~GHz in two different devices.


\section{$\textbf{Conclusions}$}
\label{sec:S4}

The results presented in this paper link pumping of electrons in a single-parameter configuration to the first charge state of a single isolated atomic potential. Although the SAT approach to QEP shows some differences from the ones based on conventional pumps, as an atom provides an isolated ground state with high charging energy, this system appears to be a promising QEP geometry. The fact that the $fe$ quantisation can be observed in two different devices, even when the gate oscillates at 1 GHz, is of interest as it was obtained at 4.2 K, without the need for a high magnetic field to increase confinement and specifically designed cycles to avoid dramatic non-adiabatic excitations. Furthermore, this result aligns with requests by the most recent theoretical proposals \cite{Wul1} and indicates that the atom potential provides \textit{naturally} a good charge pumping system.

In our work, we not only demonstrated the single atom limit of pumping in a CMOS compatible device, but also the validity of earlier theoretical predictions \cite{Blu343,FleR16291} linking high $E_{C}$'s with high $f$ of operations. The simplicity of our approach, requiring only liquid helium temperatures, should allow easy scaling to the more complicated devices recently suggested for error correction \cite{Wul1}. As there is a large research effort in the direction of the fabrication of SATs \cite{Sel206805,Roc206812,Fue242,Pla541}, progress could be rapid. In conclusion, this work demonstrates that an isolated dopant atom is not only a promising platform to build a quantum computer \cite{Pla541}, but also for QEPs.

\section*{Acknowledgments} The devices used in this work have been designed and fabricated in the AFSiD Project, see http://www.afsid.eu. The authors thanks K. Beckmann for his help during the initial phases of the experiments and A. Rossi, M. Governale and L. C. L. Hollenberg for some useful discussions during the preparation of this manuscript. G. C. Tettamanzi acknowledges financial support from the ARC-Discovery Early Career Research Award (ARC-DECRA) scheme, project title : ÒSingle Atom Based Quantum MetrologyÓ and ID: DE120100702 and from the UNSW under the GOLDSTAR Award 2013 scheme. SR acknowledges the ARC FT scheme, project ID: FT100100589.

\section*{$\textbf{Appendix}$}
\label{sec:A}

 \renewcommand\thesection{A.\arabic{section}}

 \setcounter{section}{0}

\section{$\textbf{On the non-adiabatic pumping regime of our RF data}$}
\label{sec:A1}

The RF measurements in the main manuscript are all in the quantum non adiabatic pumping regime, as the time scales (frequencies) driving our system are always $\ll$ ($\gg$) than the time scales (frequencies) associated to the dwelling of the electrons into the state. In our system the dwelling times can be extracted via the two tunnelling times of the electrons from or to the leads (i.e. via the slowest of the two). Furthermore, the DC barrier transparencies can be estimated to be more than 1600~ns (less that 625~KHz) for device $A$ and around 400~ns for device $B$. Interestingly, for device $A$, the range of the time scales (frequencies) used for the driving RF signal sits between 1~ns (1~GHz) and 40~ns (25~MHz) and therefore always $\ll$ ($\gg$) than 1600~ns (625~KHz). For device B, the system is driven at a time scale (frequency) of 1~ns (1~GHz)Ê which certainly is $\ll$ ($\gg$) than 400~ns (2.5~MHz). The tunnelling times can be extracted from the DC currents observed in the linear (low bias) regime in the following way: for device $A$ these currents can be estimated~$\lesssim$~100~$fA$ as, in the low bias regime, no signal can be distinguished from the noise back ground of the measurements in Fig.~\ref{fg:PUMPING2} of the section~\ref{sec:S2}, up to $V_{TG1}$~$=$~$V_{TG2}$~$\approx$~70~mV. Note that, in the AC experiments of device $A$, we are using an amplitude of around 0.1 V for the RF excitation and DC voltages ranging between -~0.2~V and -~0.05~V for the top gates. So we are always in the range between -~0.3~V and 0.05~V during these pumping experiments. For device $B$, the state used for pumping shows only small levels of current ($\approx$~400~fA) in the DC condition similar to the one used for pumping. 100~(400)~fA represents $\approx$~0.6~$\ast$~$10^{6}$ (2.5~$\ast$~$10^{6}$) electrons/s, and a barrier transparency on the order of $\frac{1.6}{10^{6}}$~s ($\frac{0.4}{10^{6}}$~s), hence these are in the 1.6 (0.4) $\mu$s~=~1600~(400)~ns range,  justifying our initial assumptions. However, note that there are some substantial differences between the non-adiabatic pumping limit observed in our system and the one discussed in the recent paper by Roche and co-authors \cite{Roc1581}, as in this recent work another important time scale, not available for our system, has to be taken into account: i.e. the Êcoupling or interaction between two atomic states. This because, in this recent paper also the non-adiabatic crossing of two different atomic levels is studied. Indeed, this previous work focusses in the adiabatic regime of transport for the study and characterisation (for example via Landau-Zener physics) of a two atoms system and the results that it describes are, of course, of great relevance for the field of quantum electronics based on dopants, but certainly different from the results presented in our manuscript. On another hand, also note that the extrapolations done for our system are in line with the theoretical predictions of Zimmermann \textit{et al} \cite{Zim5254} and of Kaestner \textit{et al} \cite{Kae153301} stating that, in the single parameter pumping regime situation similar to the one we are using, current will always be blocked if one of the tunnelling couplings between dot and the leads does not satisfy the condition to be É\textit{"less than a characteristic scale of $hf$"} (with $h$ being the plank constant and $f$ the frequency). As in our experiment with device $A$ quantised pumping currents are still observable at 25 $MHz$, we can, again, indirectly, extrapolate tunnel couplings $\ll$~MHz's ($\gg$~40~ns), therefore giving another strong indication of the fact that, in our AC measurements, we are always in the non-adiabatic pumping regime. Lastly, it is also important to note that, the observation of non-adiabatic quantised charge pumping plateaus up to 1 GHz at 4.2 K and without the use of complex non sinusoidal excitation, would be very difficult to be linked to an object such as a trap or a disordered QD, as none of them could have the sufficiently high charging energy ($\gtrsim$~30 meV) and the sufficiently isolated ground state, together.

\section{$\textbf{More details on the devices selection process}$}
\label{sec:A2}

\begin{figure*}
\begin{center}
\includegraphics[width=172mm]{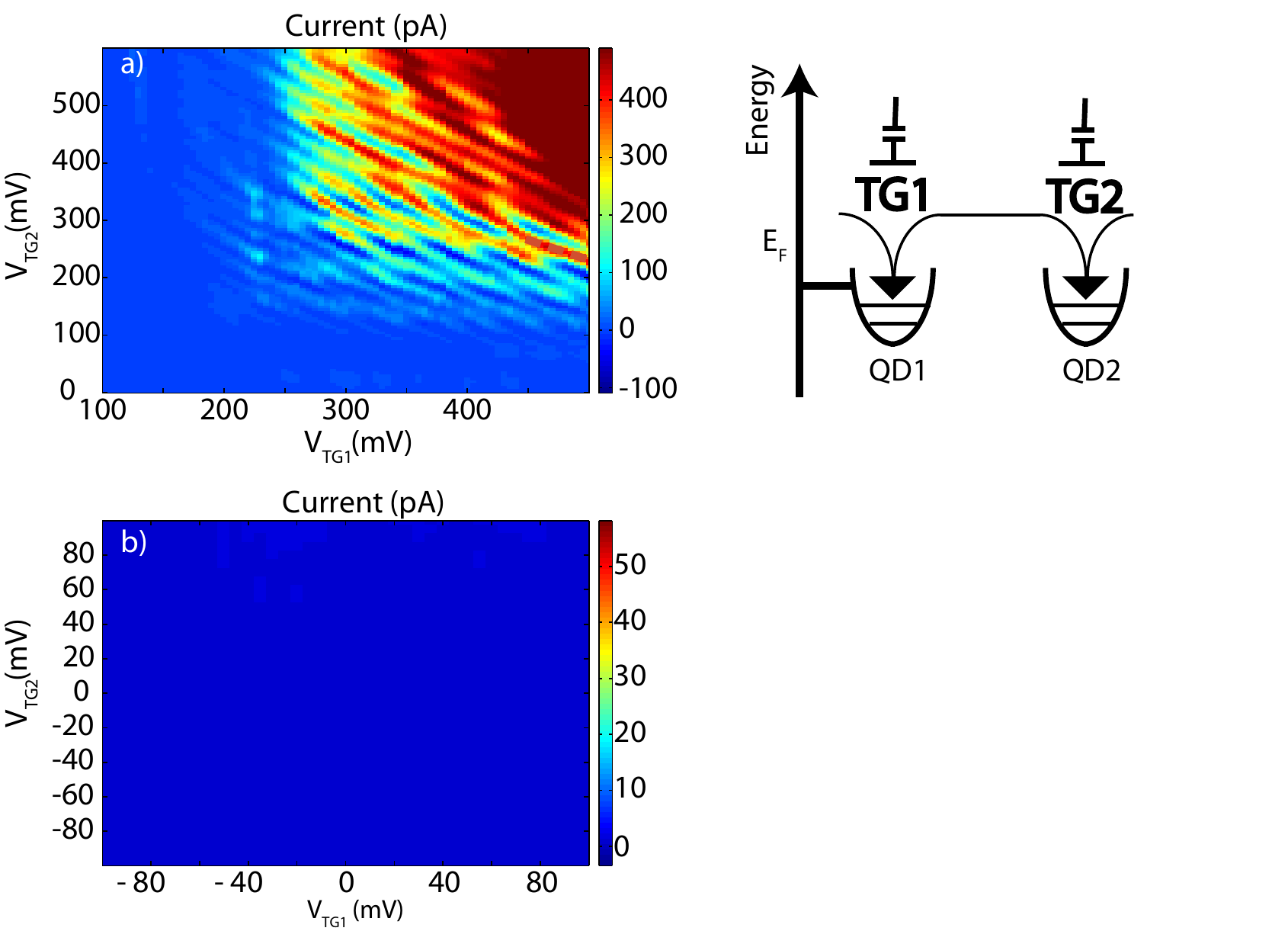}
\caption{2D current vs $V_{TG1}$ and $V_{TG2}$ stability diagram at $T$~$=$~4.2~K and $V_{SD}$ = 1~mV  for a device having similar dimensions (same $T$ and $W$ and $L_{g}$~$=$~30~nm) to devices $A$ and $B$ but $S_{gg}$~$=$~70~nm. As the transport signature is less affected by the presence of dopant-atoms in the centre of the channel, an ordered DQD \cite{Wie1} signature is observed in a) and no transport at all is observed in b). a) and b) illustrate the situations $V_{TG}$'s~$\gtrsim$~$V_{th}$ and  $V_{TG}$'s~$\lesssim$~$V_{th}$, respectively. }
\label{fg:SUPPPUMPING1tris}
\end{center}
\end{figure*}

In fact, as the data in Fig.~\ref{fg:SUPPPUMPING1tris} shows, the 2D current diagram for a device very similar to devices $A$ and $B$ but having $S_{gg}$~$=$~70~nm, e.g.~larger than~50~nm, the DQD signature is present while the dopant-atoms signature, especially in the sub-threshold regions of transport, is substantially suppressed. Indeed, in this Fig.~\ref{fg:SUPPPUMPING1tris}b), the complete lack of transport for $V_{TG1}$, $V_{TG2}$~$\lesssim$~100~mV is observed even when (not shown in the Fig.~\ref{fg:SUPPPUMPING1tris}) a large source/drain bias is applied. Furthermore, the absence of dopant related signature is also observable in the  $V_{TG1}$, $V_{TG2}$~$\gtrsim$~100~mV region of Fig.~\ref{fg:SUPPPUMPING1tris}a), i.e. in the DQD region related to the two top gates, as a regular honey-comb structure \cite{Wie1} is observed.

Lastly, if the states observed for $V_{TG1}$, $V_{TG2}$~$\lesssim$~100~mV in Fig.~\ref{fg:PUMPING2} of the main manuscript were related to a QD formed in the centre of the channel and not to dopants, a change of $S_{gg}$ of 20~nm could (slightly) influence their charging energy but would not be sufficient to shift their position of $\approx$~0.2~V. These facts are an indication of the absence of transport through isolated dopants and indeed pumping through isolated dopants has proved to not be possible in this latter case.

\section{$\textbf{Current plateaus at different frequencies}$}
\label{sec:A3}

\begin{figure*}
\begin{center}
\includegraphics[width=172 mm]{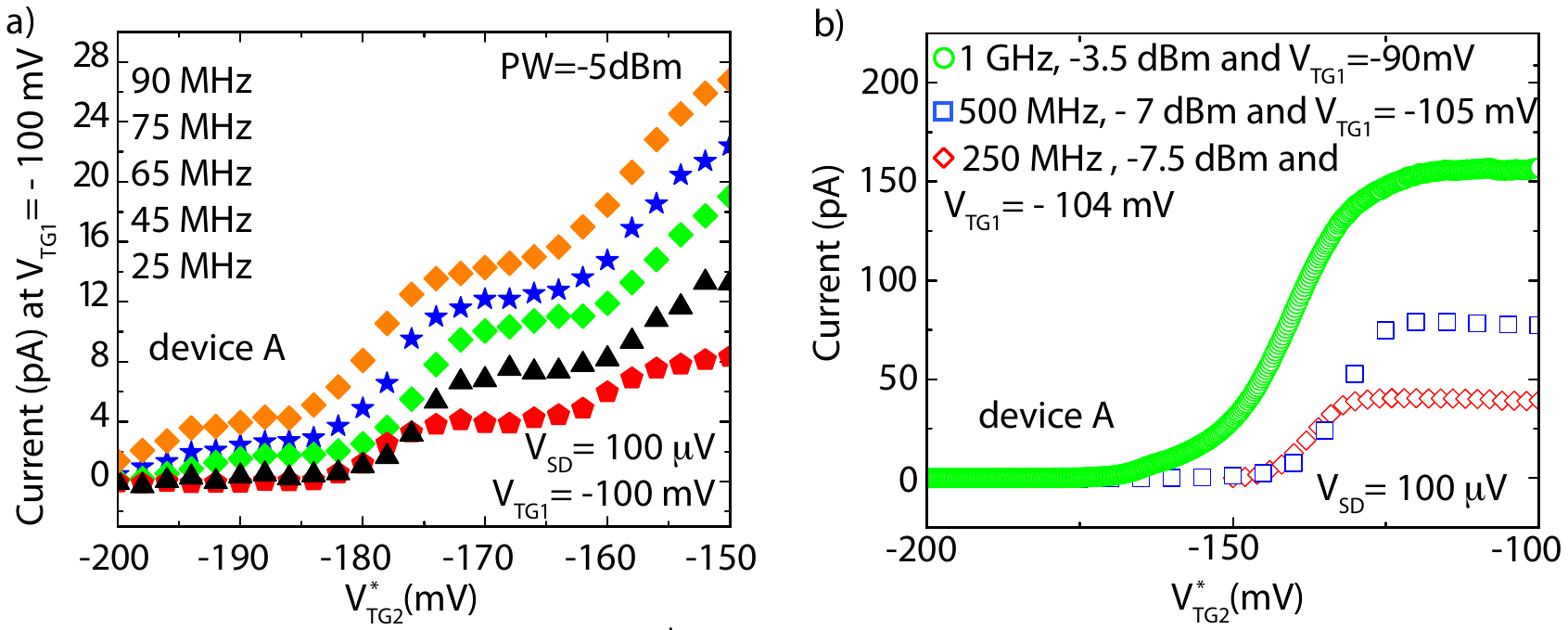}
\caption{a) Current versus $V_{TG2}$ curves for $V_{TG1}$~$=$~-~100~$mV$ and $f$~$=$~25~$MHz$,~45~$MHz$,~65~$MHz$,~75~$MHz$ and~90~$MHz$ for the $V_{RF}$ added to the DC $V_{TG2}$. b) Current versus $V_{TG2}$ curves for $V_{TG1}$~$\sim$~100~$mV$ and $f$~$=$~250~$MH$,~500~$MHz$ and~1~$GHz$ in unit of pA. These curves have been used for the extrapolation of the points in Fig.~3 c) of the main manuscript.}
\label{fg:SUPPPUMPING2}
\end{center}
\end{figure*}

In Fig.~\ref{fg:SUPPPUMPING2} the evolution of the pumping currents for device $A$ and for the different $f$ are shown.

\section{$\textbf{More details on the analysis used for device $B$}$}
\label{sec:A4}

Device $B$ was used for more accurate tests of the precision of the pumping currents in our system. To tune the barriers under the two top gates, a fixed back gate (BG) voltage value of 5.715 V was found to be an optimal value. However, due to the application of this BG, a small leakage current of a few pA's was measured both at $f$~$\approx$~1~GHz and at $f$~$\approx$~0~Hz. This leakage is exponential in gate voltage but fully independent of AC excitation. This small current was uniformly subtracted from the original data. Another consequence of the application of the back gate is that slow (few in several hours) two level charge fluctuations lead to a bi-stability of the current. As a consequence of this a different average strategy had to be implemented for this device. During 28 hours a trace was taken approximatively every 72 s and only all the traces that were unaffected by the slow charge trap were included in the average. 

\vspace{10 mm}

\centerline{{$\textbf{References}$}}


\end{document}